# Backscatter Measurements and Models for RF Sensing Applications in Cluttered Environments

Dmitry Chizhik, *Fellow, IEEE*, Jinfeng Du, *Member, IEEE,* Jakub Sapis, Reinaldo A. Valenzuela, *Fellow, IEEE*, Abhishek Adhikari, Gil Zussman, *Senior Member, IEEE*, Manuel A. Almendra, Mauricio Rodriguez, *Senior Member, IEEE*, and Rodolfo Feick, Life *Senior Member, IEEE*

*Abstract*— A statistical backscatter channel model for indoor clutter is developed for indoor RF sensing applications based on measurements. A narrowband 28 GHz sounder used a quazi-monostatic radar arrangement with an omnidirectional transmit antenna illuminating an indoor scene and a spinning horn receive antenna less than 1 m away collecting backscattered power as a function of azimuth. Median average backscatter power was found to vary over a 12 dB range, with average power generally decreasing with increasing room size. A deterministic model of average backscattered power dependent on distance to nearest wall and wall reflection coefficient reproduces observations with 4.0 dB RMS error. Distribution of power variation in azimuth around this average is reproduced within 1 dB by a random azimuth spectrum with a lognormal amplitude distribution and uniformly random phase. The model is extended to provide power distribution over both azimuth and delay (conveying range to scatterer) by combining azimuthal distribution with published results on power delay profiles in reverberant environments. The statistical model does not require a detailed room layout description, aiming to reproduce backscatter clutter statistics, as opposed to a deterministic response.

## I. INTRODUCTION

There is an increasing interest [1][2][3][4] in the joint use of communication signals for sensing, often termed Joint/Integrated Communications and Sensing (JCAS or ISAC). The new functionality is essentially that of radar, aiming at detecting and possibly characterizing some aspect of the environment. Often the ambition is to detect and localize a static or moving object, termed "target", such as a person, vehicle, robot, or UAV, in the presence of the rest of the environment, the response to which is termed "clutter". Clutter echoes are often stationary, e.g., buildings outdoors, walls and furniture indoors but may also include moving objects such as people, vehicles, as well as vegetation and various street furniture that swing with the wind. Sensing scenarios of interest include monostatic, with collocated transmitter/receiver measuring backscatter, e.g., a single base station or terminal, as well as bi-static where transmitter and receiver are separated, e.g., signal traveling from one base station to another, scattering along the way. Various use cases for joint communication and sensing are under consideration by the 3GPP [24].

Algorithms for detection, localization and, possibly, classification, of objects need to be tested in realistic scenarios, requiring representative models of both clutter and target. It is desirable for the model to be easily implementable. 3GPP TR 38.901 [5]) describes communication models which have been adopted in proposals for statistical channel models for communications and sensing[7][8][9]. While such formulations are very general, determination of properties such as strength of backscatter remains open, particularly for clutter. Clutter may be viewed as either an extended object or a collection of many objects. Backscatter from clutter is then critically dependent on the number and scattering strength of such objects, as well as multiple scattering from them. This information is not available from current 3GPP communication channel models or single object target models developed in radar.

Deterministic models, such as ray tracing, are general enough to be used in both monostatic and bi-static arrangements[20].Hybrid models such as quasi-deterministic models [18][19] (see also references therein) supplement site-specific cluster/ray generation from ray tracing with stochastic components to capture the complexity and uncertainty of the environment and the targets. Aside from issues with accuracy [6][10], such models require use of specialized, often licensed, software making standardization difficult.

Models developed for communication channels are often inapplicable to many of the sensing tasks. For example, monostatic radar measures backscatter, not of interest in communication channels where one is interested in the signal going from transmitter to a well separated receiver. Even the bi-static radar case is not well covered: commercial communications often involve propagation from base station to a terminal. Sensing might involve propagation from base to base or terminal to terminal. In all these cases new propagation models, verified by corresponding measurements are needed.

In this work we propose a measurement-based monostatic indoor backscatter model. Narrowband measurements of arriving backscatter power as a function of azimuth at 28 GHz were collected in 3 cities in 251 indoor locations in rooms of different sizes, allowing formulation of a model for indoor backscatter validated though a statistically significant data set, in environments containing both metal and dielectric materials.

Dmitry Chizhik, Jinfeng Du, Jakub Sapis and Reinaldo A. Valenzuela are with Nokia Bell Labs, Murray Hill, NJ 07974, USA (e-mail: dmitry.chizhik@nokia-bell-labs.com, jinfeng.du@nokia-bell-labs.com, jakub.sapis@nokia-bell-labs.com, reinaldo.valenzuela@nokia-bell-labs.com), Abhishek Adhikari, Gil Zussman, are with Columbia University, New York, NY, abhishek.adhikari@columbia.edu, gil.zussman@columbia.edu, Manuel A. Almendra and Mauricio Rodgriguez is with Pontificia Universidad Católica de Valparaíso, Valparaíso, Chile, (email: manuel.almendra.v@mail.pucv.cl, mauricio.rodriguez.g@pucv.cl), Rodolfo Feick is with Universidad Técnica Federico Santa María, Valparaíso, Chile (email: rodolfo.feick@usm.cl)

Based on the data, the backscatter power ratio is represented as varying over azimuth about an average, with variation statistics derived from measurements. Azimuthal power spectra measured at locations every 10 cm allowed an assessment of variability at closely spaced locations. Key results include:

- Average measured clutter backscatter power ratio was found to generally decrease with increasing room size.
- A simple theoretical model for average backscatter as a function of distance to nearest illuminated wall was derived, with 4.0 dB RMS error.
- Variation of clutter backscatter about its local average was found to be well represented as having a lognormally distributed amplitude, each with uniformly random phase. The resulting CDF of the variation statistics is within 1 dB of the observed statistics and within 20% of the observed spatial correlation of azimuthal power spectra.
- The model is extended to provide power distribution over both azimuth and delay by combining azimuthal distribution with published results on power delay profiles of reverberant scatter where transmitter and receiver are in the same room. Echo time delay conveys information on travel time, and, thus, range to scattering object.
- The statistical model does not require a detailed room layout description, aiming to reproduce backscatter clutter statistics, as opposed to a deterministic response.
- The resulting clutter model was used in combination with a simple target backscatter model for a walking person to demonstrate joint clutter-target modeling.

## II. BACKSCATTER FROM A TARGET AND CLUTTER

Received radar power of a signal scattered from an object $R_o$ meters away with effective radar cross-section (RCS) $\sigma_{scat}$ located at distance $R_s$ meters away from the transmitter is given by the radar equation [11]:

$$P_R = \frac{\lambda^2}{4\pi} \frac{1}{4\pi R_o^2} \sigma_{scat} \frac{1}{4\pi R_s^2} P_T G_T(\Omega_{inc}) G_R(\Omega_{scat}) \quad (1)$$

where $\lambda$ is carrier wavelength, $P_T$ is the transmit power, $G_T$ and $G_R$ are the antenna gains, and $\Omega_{inc}$ and $\Omega_{scat}$ are the angles of incidence and scattering, respectively.

For the quasi-monostatic antenna arrangement studied here, distances and angles from transmitter and receiver antennas to target are the same, with target backscatter power defined as:

$$P_{target} = \frac{\lambda^2 \sigma_{back} P_T G_T G_R}{(4\pi)^3 R^4} \quad (2)$$

The above equations apply when the distance $R$, and antenna gains $G_T$, $G_R$ do not change substantially over the extent of the target. This still allows for scattered power fluctuation due to fluctuation in $\sigma_{back}$, stemming from relative phase variation of scattered signals from different parts of the target.

In the case of an extended scatterer, such as a rough wall or a dense collection of distributed scatterers, (1) generalizes [11] to an integral over the surface of the extended scatterer:

$$P_{clut} = \frac{\lambda^2 P_T}{(4\pi)^3} \iint dA \, G_T(\Omega) G_R(\Omega) \frac{\gamma_{back}(\Omega)}{R^4} \quad (3)$$

With $\gamma_{back}(\Omega)$ a (unit-less) backscattering cross-section of the clutter surface per unit area.

A quantity of particular interest here is the backscatter power ratio for clutter:

$$P_{back} = \frac{P_{clut}}{P_T} \quad (4)$$

## III. MEASUREMENT DESCRIPTION

The radar measurement set-up was adopted from a narrowband channel sounder [12]. The transmitter emitted a 28 GHz CW tone at 22 dBm from an omnidirectional antenna. The receiver is a 10° (24 dBi) horn, mounted on a rotating platform allowing a full angular scan every 200 ms. The receiver records power samples at a rate of 740 samples/sec. using an onboard computer. The omnidirectional transmit antenna was placed on a lower shelf of a plastic cart (Fig. 1) to illuminate the surrounding area uniformly in all azimuthal directions. The spinning horn was placed on an upper shelf of the cart, about 0.7 m above the lower shelf. Multiple layers of absorbing foam, separated with aluminum foil attenuating the direct Tx-Rx signal path. This monostatic radar arrangement was calibrated by measuring backscattered power from a standard trihedral target with known [21] Radar Cross-Section (RCS) in an anechoic chamber. Measured backscattered power was within 1.4 dB of prediction by (2).

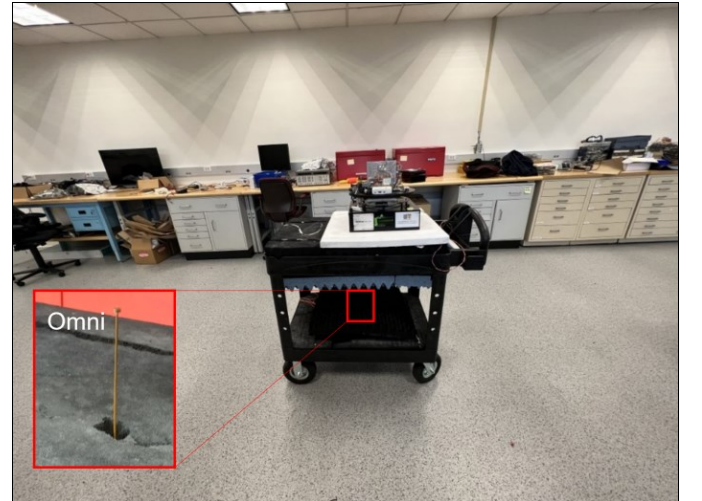

Fig. 1. 28 GHz narrowband backscatter radar arrangement, with omnidirectional Tx antenna on lower cart shelf illuminating the scene and a spinning horn receiver collecting backscatter power vs. azimuth on top cart shelf.

## IV. AVERAGE BACKSCATTERED POWER

The cart was placed in 251 locations in rooms of varying sizes, from 3x3 m offices to 20x30 m cafeteria, collecting backscatter azimuth spectra like one shown in Fig. 2. Some rooms had metal furniture and metalized windows, others wooden furniture and plain glass windows. The data was collected in 3 different building types in New York City, New

Jersey office building and a university building in Chile. Room materials varied, with some rooms containing metal furniture along the walls, others with primarily wood furniture and drywall walls. We measured 81, 128 and 42 small, medium, and large office locations, with distances to nearest illuminated wall of under 2.5, 3-5 m, and greater than 5 m, respectively. A general observation was that backscattered power ratio in any direction may be characterized statistically by variation around an average value over azimuth:

$$\langle P_{\text{back}} \rangle_\phi = \frac{\langle P_{\text{clut}} \rangle_\phi}{P_{\text{T}}} \quad (5)$$

indicated as a dashed circle in Fig. 2. The observed average backscatter value varied from location to location, with cumulative distributions, grouped by room size, plotted in Fig. 3. For display clarity, Fig. 3 shows a representative set of 73 links collected in rooms with metal furniture and /or metalized windows. Both transmit and receive antenna heights were no more than 1 m above the floor, thus primarily having metal cabinets in view in such rooms, seen in Fig. 1. It is observed that average backscatter power generally decreases as the room size increases.

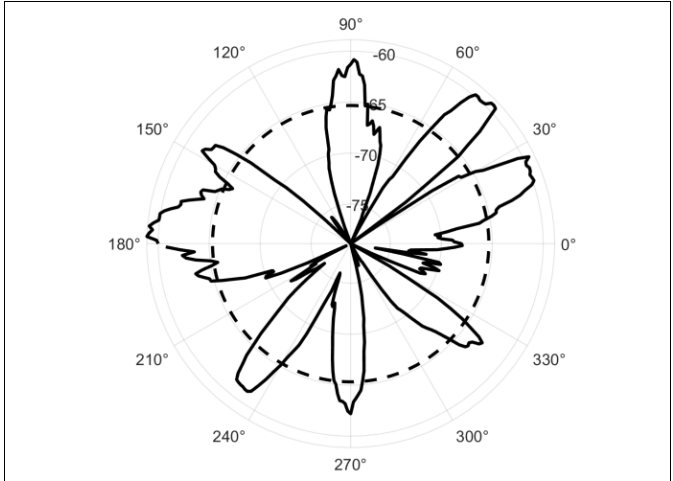

Fig. 2 Sample measured backscattered power ratio vs. azimuth. Dashed line is average backscattered power ratio (5)

In these measurements, the radar transceiver is placed near center of a room, with the visually observed clutter consisting of walls and near wall furniture, with clutter at distance $d_s$ that generally varies with direction. Viewing such clutter as an extended scatterer at distance $d_s$ from the transceiver, as illustrated in Fig. 4, we evaluate the corresponding theoretical backscatter ratio (3) for comparison against measurements.

A simple heuristic form of backscattering cross-section $\gamma_{\text{back}}(\Omega)$ is the Lambert's law [11], setting the cross section as proportional to the factor $\cos\Omega_{\text{inc}} \cos\Omega_{\text{scat}}$ projecting the scattering area $dA$ onto the directions towards the receiver and transmitter. For the monostatic case $\Omega_{\text{inc}} = \Omega_{\text{scat}} = \Omega$, thus:

$$\gamma_{\text{back}}(\Omega) = |\Gamma_{\text{wall}}|^2 \cos^2 \Omega \quad (6)$$

Where $|\Gamma_{\text{wall}}|^2$ is the ordinary Fresnel reflection coefficient for a planar surface. Lambert's law has recently been reported [13] representative of observed scatter from buildings with complex architectural features.

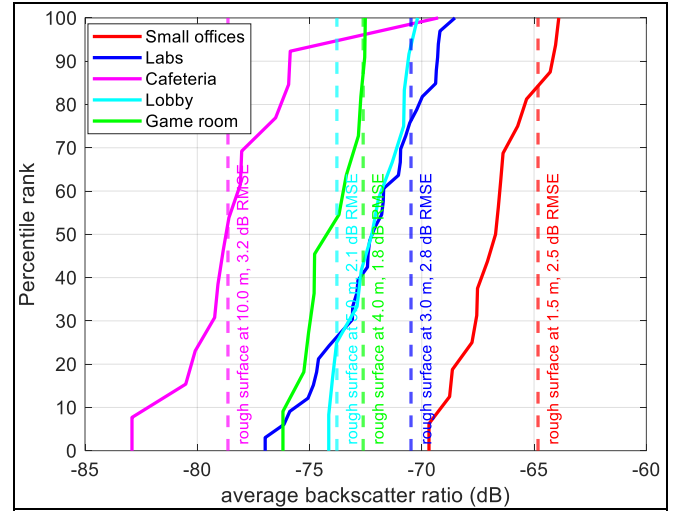

Fig. 3. Distributions of measured locally averaged backscatter power ratios (5) in rooms of different sizes. Dashed vertical lines are predictions by (9), with distance to wall as indicated

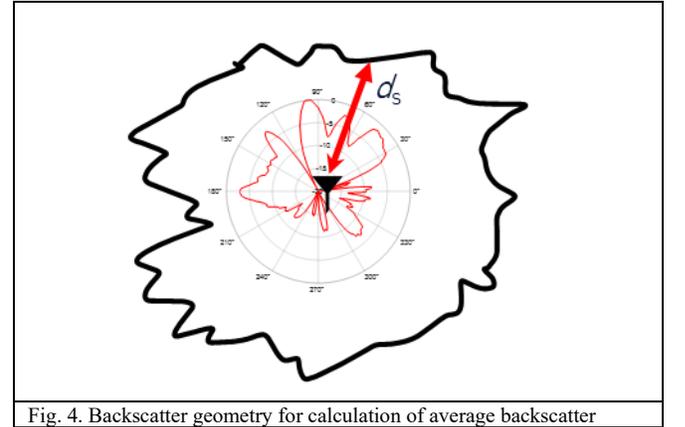

Fig. 4. Backscatter geometry for calculation of average backscatter

The scattered power integral (3) may be evaluated, particularly simply in the case of interest here, where the integrand behavior is defined by the beamspot cast on the clutter region by the directive receive antenna (10° beamwidth in these measurements):

$$G_R(\phi,\theta) = \frac{2}{\phi_{\text{RMS}} \theta_{\text{RMS}}} e^{-\phi^2/2\phi_{\text{RMS}}^2} e^{-\theta^2/2\theta_{\text{RMS}}^2} \quad (7)$$

$$\langle P_{\text{clut}} \rangle_\phi = \frac{\lambda^2 P_T |\Gamma_{\text{wall}}|^2}{(4\pi)^3} \iint dA \ G_R(\phi,\theta) G_T(\phi,\theta) \frac{\cos^2 \Omega}{R^4}$$

$$= \frac{\lambda^2 P_T |\Gamma_{\text{wall}}|^2 2 G_T d_s^2}{(4\pi)^3 \phi_{\text{RMS}} \theta_{\text{RMS}} d_s^4} \int_{-\pi/2}^{\pi/2} d\theta \ e^{-\theta^2/2\theta_{\text{RMS}}^2} \int_{-\pi}^{\pi} d\phi e^{-\phi^2/2\phi_{\text{RMS}}^2} \quad (8)$$

$$\approx P_T G_T |\Gamma_{\text{wall}}|^2 \left(\frac{\lambda}{4\pi d_s}\right)^2$$

Where differential area $dA = d_s^2 d\theta d\phi$, while $\cos^2\Omega \approx 1$, $R \approx d_s$ for region where the antenna pattern $e^{-\theta^2/2\theta_{RMS}^2}e^{-\phi^2/2\phi_{RMS}^2}$ is significant. The transmit antenna is assumed to have a much broader beamwidth than the receive antenna, aimed in the same direction. In measurements done here an omni antenna was used in the transmitter, with $G_T=1$. In evaluating (8) it was assumed that the integrand effective angular support is defined primarily by the directivity of the (receive) antenna, as appropriate for the 10° antenna used in measurements in this work. Under these conditions the average backscattered power $P_{clut}$ is independent of the antenna patterns.

Average backscattered power ratio can thus be defined in terms of environment quantities alone:

$$\langle P_{back}\rangle = \frac{\langle P_{clut}\rangle_\phi}{P_T} = |\Gamma_{wall}|^2\left(\frac{\lambda}{4\pi d_s}\right)^2 \qquad (9)$$

TABLE I.
BACKSCATTER POWER RATIO IN DIFFERENT SIZE ROOMS
Measured and predicted by (9), at distance $d_s$ to nearest wall.

| Data Set | # of links | Room dimensions | $d_s$ (m) | Measured median average backscatter ratio (dB) | Model (9) RMS error (dB) |
|---|---|---|---|---|---|
| 8 offices, metal furniture | 16 | 3m × 3m | 1.5 | -66.7 | 2.5 |
| 7 Labs, metal furniture | 33 | 20m × 6m | 3.0 | -72.1 | 2.8 |
| Cafeteria, metalized windows, counters | 13 | 30m×20m | 10.0 | -78.6 | 3.2 |
| Lobby, metalized windows | 12 | 30m×20m | 5.0 | -72.1 | 2.1 |
| Game room, metalized windows | 11 | 25m×15m | 4.0 | -73.7 | 1.8 |
| Conference Room | 26 | 7m×5m | 2.5 | -72.6 | 3.2 |
| Gym | 17 | 30m×17m | 9.0 | -78.7 | 5.4 |
| Lab 1 | 15 | 15m × 5m | 2.5 | -71.1 | 4.7 |
| Lab 2 | 10 | 15m × 5m | 2.5 | -71.1 | 4.3 |
| Office 1 | 22 | 3.5m×2.7m | 1.0 | -69.3 | 2.4 |
| Office 2 | 17 | 4m × 3m | 1.5 | -71.8 | 2.7 |
| Study hall | 22 | 15m × 9m | 4.5 | -71.7 | 7.7 |
| Café | 15 | 14m×10m | 3.0 | -71.4 | 5.2 |
| Carleton Hall | 22 | 25m×12m | 3.0 | -74.8 | 2.2 |
| **Overall** | **251** | | | | **4.0** |

Average backscatter ratio predicted by (9) is evaluated for different room sizes and compared against measurements in Fig. 3. Rooms with steel furniture dominating the field of view of antennas (as in Fig. 1), and metalized windows, were assigned $|\Gamma_{wall}|^2 = 1$ in (9), while rooms with wooden furniture were assigned $|\Gamma_{wall}|^2 = 0.25$, corresponding to the power reflection coefficient from air-dielectric interface with relative dielectric constant of 3 [22] (representative of dry wall/wood), averaged over a uniform distribution of incidence angles over 0-90°. Distance to clutter $d_s$ was set to distance to nearest wall used in measurements, typically $d_s$=half the smallest dimension of the room. Formula (9) prediction is indicated as corresponding vertical lines in Fig. 3 for various room sizes. Room dimensions and measured backscatter power for all data sets is tabulated in Table I along with RMSE error and is found to predict measured average backscatter ratio with 4.0 dB RMS error, notable considering the medians of the observed distributions span some 12 dB. Thesole parameters in (9) are the distance $d_s$ to nearest illuminated wall and material-dependent reflection coefficient $|\Gamma_{wall}|^2$ of 1 or 0.25, for metal/dielectric surroundings, respectively.

## V. OBSERVED AZIMUTH VARIATION STATISTICS

As illustrated in Fig. 2, measured backscattered power exhibits variation with azimuth around its local average. The distribution of observed power variations (9) is plotted as a solid line in Fig. 5.

Spatial variation of observed azimuth spectra was examined in a lab, with locations along a 1 m long line segment every 10 cm, for a total of 11 azimuthal spectra. Correlation coefficient $\rho(d_2 - d_1)$ of azimuthal spectra was computed for different separations $d_2 - d_1$ using measured spectra $P(\phi, d_n)$ at locations $d_n$ as:

$$\rho(d_2-d_1) = \langle p(\phi,d_2)p(\phi,d_1)\rangle,$$
$$p(\phi,d_n) \equiv \frac{P(\phi,d_n)-\langle P(\phi,d_n)\rangle}{\langle (P(\phi,d_n)-\langle P(\phi,d_n)\rangle)^2\rangle^{1/2}}, \quad n=1,2 \quad (10)$$

Correlation coefficient computed from data is plotted in Fig 6 as a function of location separation $|d_2 - d_1|$. The correlation coefficient is seen to drop to about 0.3 for location separation as small as 0.1 m, indicating rapid variation in azimuthal spectra even with small displacement.

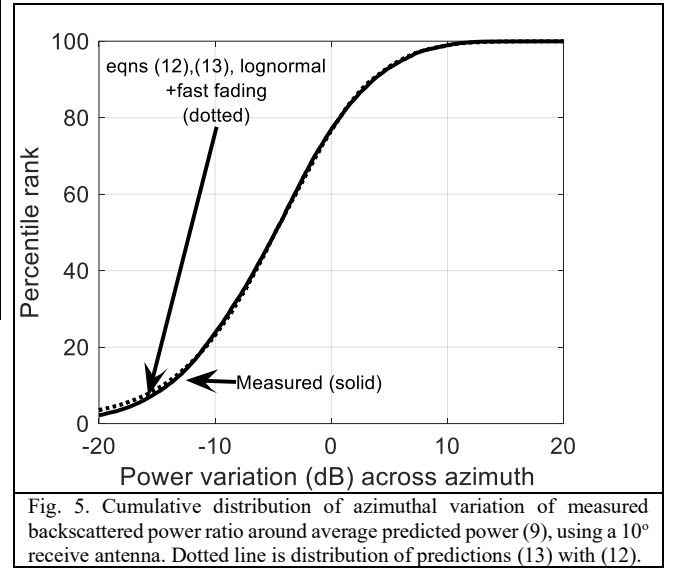

Fig. 5. Cumulative distribution of azimuthal variation of measured backscattered power ratio around average predicted power (9), using a 10° receive antenna. Dotted line is distribution of predictions (13) with (12).

Autocorrelation of azimuth spectra was computed from measured azimuth spectra, normalized as in (10):

$$\rho(\phi_2-\phi_1) = \langle p(\phi_1)p(\phi_2)\rangle, \qquad (11)$$

The averaging in is over all measured locations. The autocorrelation is plotted in Fig. 7 as a function of azimuth shift $\phi_2 - \phi_1$, superimposed on the antenna azimuth pattern (dashed) measured in a chamber. The main lobe of measured autocorrelation of backscattered azimuth spectra is nearly identical to the nominal antenna pattern, indicating that the correlation of backscatter in azimuth scale is much narrower than the antenna beamwidth.

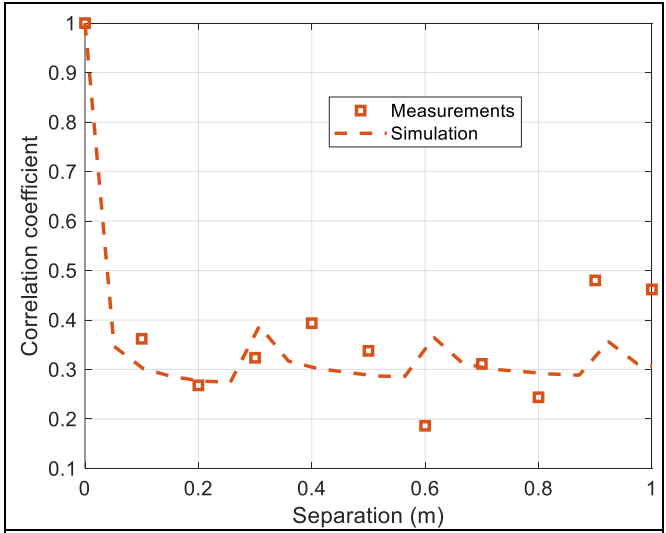

Fig. 6. Correlation of azimuthal spectra (10) at locations at different separations: Measured values are squares, dashed line is simulation evaluating (10) using predicted spectra (13) with (12)

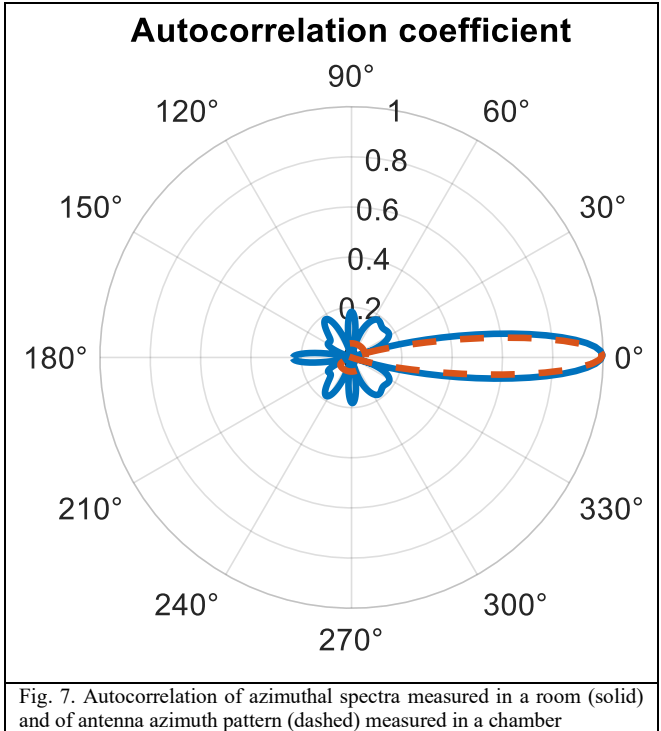

Fig. 7. Autocorrelation of azimuthal spectra measured in a room (solid) and of antenna azimuth pattern (dashed) measured in a chamber

## VI. CLUTTER BACKSCATTER MODEL

Measurements described above are used to define a model for the monostatic arrangement, with backscattered plane waves arriving at radar location **r** from azimuth $\phi$ with complex amplitude $h_{\text{chan}}(\mathbf{r}, \phi)$ given by:

$$h_{\text{chan}}(\mathbf{r}, \phi) = \sqrt{P_0(d_s)} \sqrt{10^{P_v/10}} \sqrt{10^{P(\phi)/10}} e^{i\varphi} e^{i2\mathbf{k}\cdot\mathbf{r}},$$

$$P_0(d_s) = |\Gamma_{\text{wall}}|^2 \left(\frac{\lambda}{4\pi d_s}\right)^2, \quad \text{Average backscatter power}$$

$$P_v \sim N(\mu_v, \sigma_v), \ \sigma_v = 4.0 \text{ dB}, \ \mu_v = -10\log_{10} e^{(0.1*\ln(10)*\sigma_v)^2/2}$$

$$P(\phi) \sim N(\mu, \sigma), \ \sigma = 7.0 \text{ dB}, \ \mu = -10\log_{10} e^{(0.1*\ln(10)*\sigma)^2/2}, \text{i.i.d.}$$

$$\langle P(\phi_1)P(\phi_2)\rangle = e^{-\phi^2/2\phi_{\text{rms}}^2}, \phi_{\text{rms}} = 1^o$$

$$\varphi \sim U(0, 2\pi), \mathbf{k} = (k\cos\phi, k\sin\phi) \quad \text{small-scale fading}$$

(12)

The above model includes theoretical formula (9) for average backscatter $P_0(d_s)$ at distance $d_s$ to clutter, its variation $P_v$,(dB), azimuthal power variation $P(\phi)$ (dB) relative to local average and small-scale fading expected in scattering environments, represented by plane wave arrivals each with random phase $\varphi$ and spatial variation $e^{i2\mathbf{k}\cdot\mathbf{r}}$, with a factor of 2 accounting for 2-way phase delay between transceiver and extended clutter. The random phase for each plane wave arrival is necessary to emulate small scale spatial fading, while the factor $e^{i2\mathbf{k}\cdot\mathbf{r}}$ allows for spatially consistent channel response should one move the monostatic radar station or add another antenna in proximity. The random azimuthal spectra $P(\phi)$ in (12) are modeled as having Gaussian form correlation $\langle P(\phi_1)P(\phi_2)\rangle = e^{-\phi^2/2\phi_{\text{rms}}^2}$, with characteristic scale $\phi_{\text{rms}} = 1^o$.

A direct way to simulate a correlated Gaussian process is to convolve a white Gaussian process with a corresponding filter. Normal distributions (in dB) $P_v$ and $P(\phi)$ are characterized by corresponding standard deviations $\sigma_v$ and $\sigma$, determined from observed distributions. Their mean values, $\mu_v = -10\log_{10} e^{(0.1*\ln(10)*\sigma_v)^2/2}$, $\mu = -10\log_{10} e^{(0.1*\ln(10)*\sigma)^2/2}$, correspondingly, follow to enforce $\langle 10^{P_v/10}\rangle = \langle 10^{P(\phi)/10}\rangle = 1$, since these are deviations from local average power.

Backscattered signal received by a transceiver with transmit and receive antennas aimed in directions $\phi_T$ and $\phi_R$, respectively are obtained by integrating arrival spectrum (12) over all angles, weighted by the product of transmit and receive antenna field patterns, $f_T(\phi)$ and $f_R(\phi)$:

$$y_R(\mathbf{r}, \phi_R, \phi_T) = \sqrt{P_T} \int_0^{2\pi} d\phi \ h_{\text{chan}}(\mathbf{r}, \phi) f_R(\phi - \phi_R) f_T(\phi - \phi_T) \quad (13)$$

Relative transceiver location **r** is relevant only if channels at multiple nearby (within ~ 1m) transceiver locations are considered. The antenna patterns are normalized for unit total power:

$$\int_{-\pi}^{\pi} d\phi |f_R(\phi)|^2 = \int_{-\pi}^{\pi} d\phi |f_T(\phi)|^2 = 1 \quad (14)$$

Patterns obtained by spinning directional antenna are then used for evaluations of (13) at different angles $\phi_R, \phi_T$, which constitutes a circular convolution of the channel response with antenna field patterns. The result is illustrated in Fig. 8, showing sample measured patterns on the left and sample simulated patterns on the right. Displayed azimuth spectra are individual instantiations of the random process, whose accuracy against measurements is judged based on statistical correspondence shown in Figs. 5,6,7.

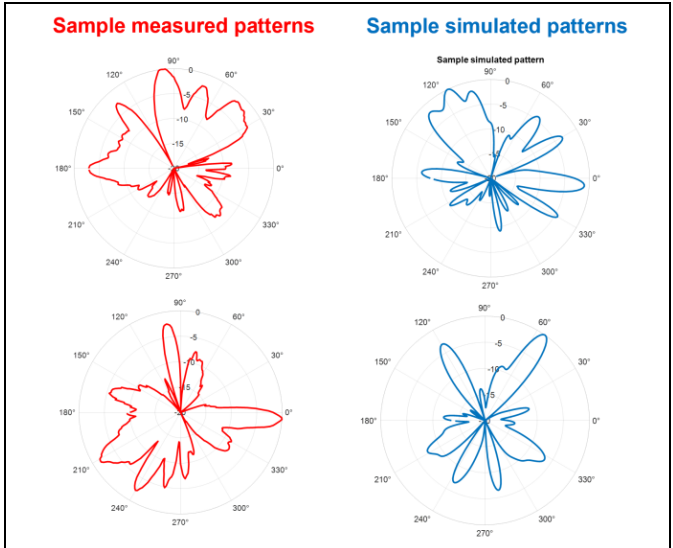

Fig. 8. Sample instantiations of backscatter azimuth spectra: measured are on left, simulated using (13) on right.

Azimuthal variation of the angular spectra generated using (12) and (13) is found to be within 1 dB of the observed distribution of backscatter power variation, as shown in Fig. 5.

Azimuthal spectrum correlation (10) of simulated spinning antenna responses (13), plotted in Fig. 6 as a function of antenna separation, is seen to correspond well to observed spectrum correlations. Autocorrelation of the simulated spectra plotted in Fig. 7 closely corresponds to the measured result, with main lobe width within 1 deg of measurements.

VII. DELAY-AZIMUTH CLUTTER BACKSCATTER MODEL

The clutter backscatter model presented in VI is now extended to include backscattered power distribution in delay as well.

Distribution of power in the channel as a function of delay has been modeled in [16][17] as a superposition of the direct arrival and a reverberant power distributed over delay. In monostatic radar the direct arrival (termed TX/RX leakage) is not of interest and effort is made to diminish its impact, usually through a combination of Tx/Rx isolation and time-gating. The reverberant power delay profile is represented in [17] as having an average delay envelope as decaying exponentially, with a close correspondence to measurements, including the simple version of the Saleh-Valenzuela model [23].

$$P_d(\tau) = e^{-(\tau - 2d_s/c)/T_{rev}} u(\tau - 2d_s/c), \quad (15)$$

Where $u(\tau - 2d_s/c)$ is the step function and peak power is set to unity. The characteristic time $T_{rev}$ is related [17] to room volume $V$, total wall area $S$ and wall absorption coefficient, with a representative value $T_{rev}=10^{-8}$ sec. The reverberation power sequence is taken as starting at $\tau = 2d_s/c$, with $2d_s$ round-trip distance to nearest illuminated wall, typically half the room size.

Joint distribution of power in azimuth and delay can be generated by combining (12) and (15), for a single transceiver (thus relative **r**=0):

$$h_{chan}(\tau, \phi) = \sqrt{P_0(\tau, d_s)} \sqrt{10^{P_v/10}} \sqrt{10^{P(\tau,\phi)/10}} e^{i\varphi},$$

$$P_0(\tau, d_s) = |\Gamma_{wall}|^2 \left(\frac{\lambda}{4\pi d_s}\right)^2 e^{-(\tau - 2d_s/c)/T_{rev}} u(\tau - 2d_s/c)$$

$$P_v \sim N(\mu_v, \sigma_v), \ \sigma_v = 4.0 \text{ dB}, \ \mu_v = -10\log_{10} e^{(0.1*\ln(10)*\sigma_v)^2/2}$$

$$P(\tau, \phi) \sim N(\mu, \sigma), \ \sigma = 7.0 \text{ dB}, \ \mu = -10\log_{10} e^{(0.1*\ln(10)*\sigma)^2/2}$$

$$\langle P(\phi_1)P(\phi_2) \rangle = e^{-\phi^2/2\phi_{rms}^2}, \phi_{rms} = 1^o$$

$$\varphi \sim U(0, 2\pi), \text{random phase for small-scale fading}$$

$$(16)$$

Here the random distribution of backscattered arrivals in delay and azimuth is described as a lognormally distributed "clutter map" $P(\tau, \phi)$.

The channel response collected with a probing waveform $x(t)$ of bandwidth $B$, using transmit and receive antenna field patterns, $f_T(\phi)$ and $f_R(\phi)$ is obtained by convolving in angle channel impulse response (16) with corresponding antenna patterns as in (13) as well as convolving in delay with $x(t)$, normalized to unit total power:

$$\int_0^\infty |x(t)|^2 = 1 \quad (17)$$

The resulting power delay profile instantiation is shown in Fig. 9, where the channel response (16) was convolved with a Hamming window 1 ns duration, here corresponding to 1/*BW*.

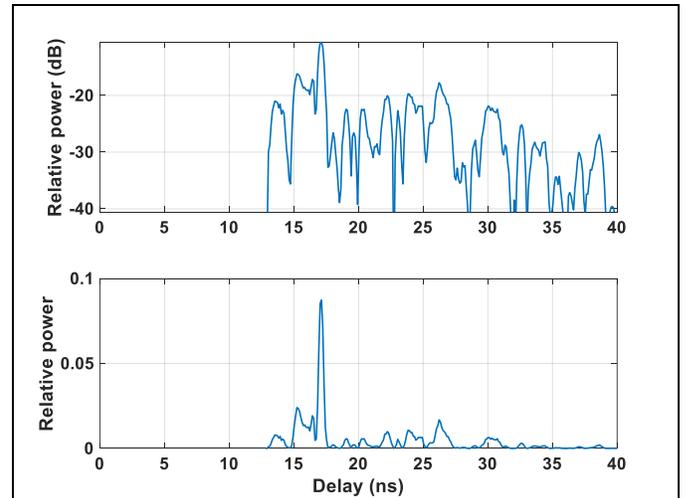

Fig. 9. Sample backscatter power delay profiles for 1 GHz bandwidth, with relative power in dB (top) and in linear units (below).

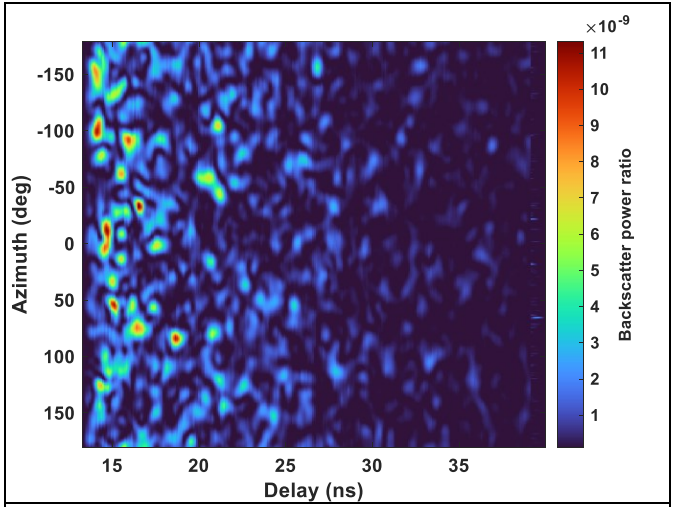

Fig. 10. Sample backscatter power vs. azimuth and delay, for 1 GHz bandwidth, 10° receive antenna and omnidirectional transmit antenna

The corresponding power distribution in azimuth and delay, obtained for 1 GHz bandwidth, 10° receive antenna and omnidirectional transmit antenna is illustrated in Fig 10. Strong arrivals appear as bright spots, usually at shorter delays, with power decaying exponentially with delay, as prescribed by (15) and (16). Model accuracy in reproducing observed azimuthal statistics is discussed in Sec. VI. Key parameters, as described above, include room dimensions, dominant materials (metal or dielectric), observed standard deviation of backscatter variabilities, $\sigma_v$ and $\sigma$ and reverberation time $T_{rev}$. The statistical model does not require a detailed room layout description, aiming to reproduce backscatter clutter statistics, as opposed to a deterministic response.

## VIII. TARGET BACKSCATTER MODEL

This section illustrates generation of total monostatic radar response in an indoor environment, containing both clutter, as modeled above, and a person walking across the room. Total radar response for a spinning antenna would be a sum of the clutter backscatter, given by (13) indoors and target response (2) (neglecting multiple scatter between clutter and target and assuming LOS to target). An important special case is that of a target moving against static background clutter. Moving target radar return is a generalization of (2), where the return power is now fluctuating with time, through time-dependent range $R(t)$, angle $\phi(t)$ and backscatter cross-section $\sigma_{back}(t)$:

$$P_{target}(t) = \frac{\lambda^2 \sigma_{back}(t) P_T G_T(\phi(t)) G_R(\phi(t))}{(4\pi)^3 R(t)^4} \quad (18)$$

Notably the backscattering cross-section $\sigma_{back}(t)$ is generally fluctuating in time both due to changes in target orientation as well as relative motion of different target components, sometimes termed "micro-Doppler", particularly relevant in targets with time-varying shape, such as people in motion. A simple way to represent such fluctuations is through Swerling models [14], developed in traditional radar applications:

$$\sigma_{back}(t) = \sigma_o |\xi(t)|^2,$$
$$\sigma_o = A_{geom} \gamma \quad (19)$$
$$\xi(t) \sim CN(0,1)$$

Where the average target cross-section $\sigma_o$ is a product of target geometric cross sectional $A_{geom}$ and a loss factor $\gamma$ representing, for example, reflection loss from a human body. As an example, $\sigma_o$ has been reported (in decibels) as -8 dB/m² [15]. The scattered field fluctuation is assumed to follow a complex Gaussian distribution, whose power is, thus, exponentially distributed, leading to the Swerling I model. This is a common assumption for arrivals consisting of many multipaths (from different target parts) with time-varying relative phases. The rate of target backscatter time fluctuation is determined by relative motion of body parts and can be obtained from measured Doppler spectrum studies. Here it is estimated using general considerations: for assumed relative limb speed of $v_{target}$= 0.1 m/s, the maximum Doppler frequency $fv_{target}/c \sim 10$ Hz at $f$ = 28 GHz, with coherence time scale of ~ 0.1 sec. The fluctuation $\xi(t)$ in (19) is here simulated by convolving a white noise sequence with a filter (chosen as Gaussian shape) with characteristic width of 0.1 sec. Alternative target models may involve other distributions, including lognormal, as found effective for clutter backscatter in this work.

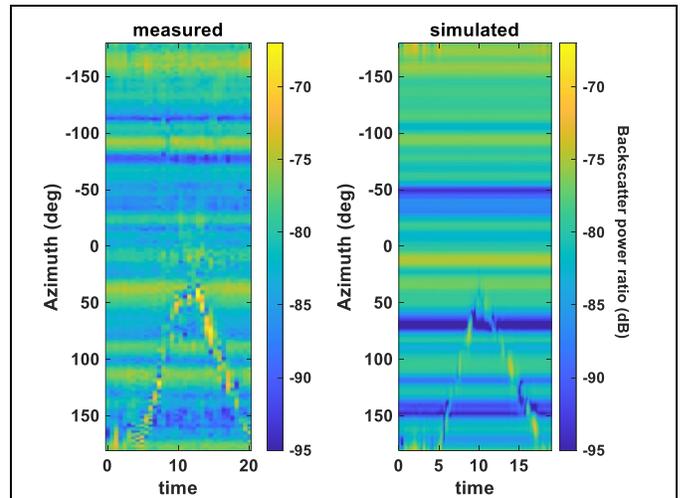

Fig. 11. Measured (left) and simulated (right) instantiation of backscatter power ratio in azimuth-time plane, with a pedestrian walking to room center and back.

Measured and simulated backscatter power ratio are shown side-by-side in Fig. 11. This measurement was collected with the radar transceiver placed in a lab, with a person walking from the side of the lab to the center and back. The trajectory of the person in azimuth-time domain is seen as a triangular shaped disturbance in the lower part of the plots of Fig. 11. The simulated clutter and target backscatter are an instantiation of the statistical model described in Sections VI, (13) with (12), and VIII, (18) with (19), respectively. Observed temporal fluctuation of clutter returns at time-angles away from the target are due to human observers on the sidelines of the room as well as secondary echoes from target and clutter. Both effects are important to characterize in future studies.

IX. CONCLUSIONS

A simple statistical monostatic radar channel model for indoor clutter is developed for indoor applications, requiring only a handful of parameters, including room dimensions, materials and measurement-derived standard deviation of azimuthal power variation. A narrowband 28 GHz sounder was used in a quasi-monostatic antenna arrangement with an omnidirectional transmit antenna illuminating the scene and a spinning horn receive antenna collecting backscattered power as a function of azimuth. Average backscatter power ratio was found to vary over a 12 dB range, with average power generally decreasing with increasing room size. A deterministic model of average backscattered power dependent only on average distance to clutter reproduces observations with 4.0 dB RMS error. Distribution of power variation in azimuth around this average is reproduced within 1 dB by a random azimuth spectrum with a lognormal amplitude distribution and random phase. The model is extended to provide power distribution over both azimuth and delay by combining azimuthal distribution with published results on power delay profiles in reverberant environments, defined by reverberation time parameter, also related to room dimensions and materials. The statistical model does not require a detailed room layout description, aiming to reproduce backscatter clutter statistics, as opposed to a deterministic response.

The statistical model of clutter and a fluctuating target is demonstrated and compared against observations for a pedestrian walking across the scene with cluttered background showing promising correspondence.

X. ACKNOWLEDGEMENTS

Thanks David Chen (Stuyvesant High School) and Timothy Wang (University of Michigan) for assistance in measurements. A. Adhikari and G. Zussman wish to acknowledge the support through NSF grants EEC- 2133516, CNS-2148128, AST-2232. M. A. Almendra, M. Rodriguez, and R. Feick wish to acknowledge the support received from the Chilean Research Agency ANID, through research grants ANID FONDECYT 1211368, and ANID PIA/APOYO AFB220004; and the project VRIEA-PUCV 039.367/2023.